\documentclass[twocolumn]{aastex631}

\usepackage{xspace}
\usepackage{tablefootnote}

\usepackage{lineno}

\newcommand{\OII}{[\ion{O}{2}]\xspace}

\newcommand{\oii}{[O\,{\sc ii}]\xspace}

\shorttitle{Dark Energy Explorers}
\shortauthors{Lindsay R. House}

\begin{document}

\title{Participatory Science and Machine Learning Applied to Millions of Sources\\ in the Hobby-Eberly Telescope Dark Energy Experiment}

\author[0000-0002-1496-6514]{Lindsay R. House}
\altaffiliation{NSF Graduate Research Fellow}
\affiliation{Department of Astronomy, The University of Texas at Austin, 2515 Speedway Boulevard, Austin, TX 78712, USA}
\correspondingauthor{Lindsay R. House}
\email{lindsay.r.house@gmail.com}

\author[0000-0002-8433-8185]{Karl Gebhardt}
\affiliation{Department of Astronomy, The University of Texas at Austin, 2515 Speedway Boulevard, Austin, TX 78712, USA}

\author[0000-0003-0792-5877]{Keely Finkelstein}
\affiliation{Department of Astronomy, The University of Texas at Austin, 2515 Speedway Boulevard, Austin, TX 78712, USA}

\author[0000-0002-2307-0146]{Erin Mentuch Cooper}
\affiliation{Department of Astronomy, The University of Texas at Austin, 2515 Speedway Boulevard, Austin, TX 78712, USA}
\affiliation{McDonald Observatory, The University of Texas at Austin, 2515 Speedway Boulevard, Austin, TX 78712, USA}

\author[0000-0002-8925-9769]{Dustin Davis}
\affiliation{Department of Astronomy, The University of Texas at Austin, 2515 Speedway Boulevard, Austin, TX 78712, USA}

\author[0000-0003-2575-0652]{Daniel J. Farrow}
\affiliation{Centre of Excellence for Data Science, Artificial Intelligence \& Modelling (DAIM),\\ University of Hull, Cottingham Road, Hull, HU6 7RX, UK} \affiliation{E. A. Milne Centre for Astrophysics
University of Hull, Cottingham Road, Hull, HU6 7RX, UK}

\author[0000-0001-7240-7449]{Donald P. Schneider}
\affiliation{Department of Astronomy and Astrophysics, The Pennsylvania State University, University Park, PA 16802, USA}
\affiliation{Institute for Gravitation and the Cosmos, The Pennsylvania State University, University Park, PA 16802, USA}

\begin{abstract}

We are merging a large participatory science effort with machine learning to enhance the Hobby-Eberly Telescope Dark Energy Experiment (HETDEX). Our overall goal is to remove false positives, allowing us to use lower signal-to-noise data and sources with low goodness-of-fit. With six million classifications through \textit{Dark Energy Explorers}, we can confidently determine if a source is not real at over 94\% confidence level when classified by at least ten individuals; this confidence level increases for higher signal-to-noise sources. To date, we have only been able to apply this direct analysis to 190,000 sources. The full sample of HETDEX will contain around 2-3M sources, including nearby galaxies (\OII emitters), distant galaxies (Lyman-$\alpha$ emitters or LAEs), false positives, and contamination from instrument issues. We can accommodate this tenfold increase by using machine learning with visually-vetted samples from \textit{Dark Energy Explorers}. We have already increased by over ten-fold in number of sources that have been visually vetted from our previous pilot study where we only had 14,000 visually vetted LAE candidates. This paper expands on the previous work increasing the visually-vetted sample from 14,000 to 190,000. In addition, using our currently visually-vetted sample, we generate a real or false positive classification for the full candidate sample of 1.2 million LAEs. We currently have approximately 17,000 volunteers from 159 countries around the world. Thus, we are applying participatory or citizen scientist analysis to our full HETDEX dataset, creating a free educational opportunity that requires no prior technical knowledge.

\end{abstract}

\keywords{Dark Energy, Machine Learning, Citizen Science, Participatory Science}

\section{Introduction}
\label{sec:intro}
The continually increasing size of the astronomical datasets requires leveraging new analysis techniques in order to handle these efficiently and extract robust scientific results. While use of machine learning (ML) has been available for decades, its use is now an essential component within the field \citep{ML_overview_astro}. Machine learning is a subfield of artificial intelligence where algorithms are used to recognize patterns, make predictions, and even apply these results to new applications \citep{zawacki-richter_systematic_2019, Torney}. 

One of the primary issues with machine learning is determining the accuracy for a given application and interpreting results. There are multiple ways to assign accuracy, including visual vetting on a subset of the sample. Human visual vetting, even for verification, quickly becomes intractable as the datasets increase in size. In fact, many of the data samples have surpassed the ability to use visual vetting within a given collaboration due to the limited number of eyes available.

Many large astronomical experiments have already paved the way in using human classification to reach science goals. The collaboration between scientists and the public is known as participatory science, also called citizen science \citep{cs_ps, cs_ps2}. There are multiple ways to include these participants within the various pipelines. A large survey that have done so is the Cosmic Assembly Near-infrared Deep Extragalactic Legacy Survey (CANDELS), and the complementary participatory science project, Galaxy Zoo \citep{Masters_2019, simmons_galaxy_2016}. In addition, the Laser Interferometer Gravitational Wave Observatory (LIGO)  utilizes the participatory science project, Gravity Spy \citep{zevin_gravity_2024}.

The Hobby-Eberly Telescope Dark Energy Experiment (HETDEX) is designed to study the expansion rate of the universe at $1.9 < z < 3.5$ with an accuracy comparable to even the best low-$z$ experiments \citep{Gebhardt21, Hill_2021}. To date, this multi-year program has already generated nearly one billion spectra, and one trillion resolution elements. The shear size of this dataset necessitates robust statistical techniques to identify false positives, \oii contamination, and artifacts. Even with the most detailed analysis, we have not yet reached our design specifications for false positive rate \citep{erincooper}. Therefore we have created and use the participatory science project \textit{Dark Energy Explorers}\footnote{\url{https://www.zooniverse.org/projects/erinmc/dark-energy-explorers}} to improve HETDEX with machine learning techniques. 

In \cite{House_2023}, we provide the first use of ML to the HETDEX database. There, we find that the \textit{Dark Energy Explorers} results are accurate to over 95\% with the ability to find false positives. This work only considered 10\% of the full HETDEX database. Given the success we had in using ML and \textit{Dark Energy Explorers}, our goal is to apply to the full database, where we need to classify about 10 million spectra.

The current database for HETDEX has about 1.2 million Lyman-$\alpha$ emitters (LAEs) and about 1 million OII emitting galaxies. These numbers come from a down selection of spectra based on the signal-to-noise (SN) and goodness-of-fit of the emission line. The initial sample is about 10 million spectra. We realize that our selection, as it is, fails to exclude all false positives and fails to include all true positives. The more sources we can robustly identify increases our cosmological constraints on the expansion of the universe by about the square root of the numbers \citep{Davis2022, Gebhardt21, erincooper, Farrow2021, Leung2017}. Similarly, the false positive rate makes our accuracy worse. Therefore, the larger the total number of LAE sources and the less false positives due to noise (FP) and \OII contamination will allow us to more accurately measure the expansion rate of the universe. 

\begin{figure*}[ptb]
   \begin{center}
         \includegraphics[scale=0.4]{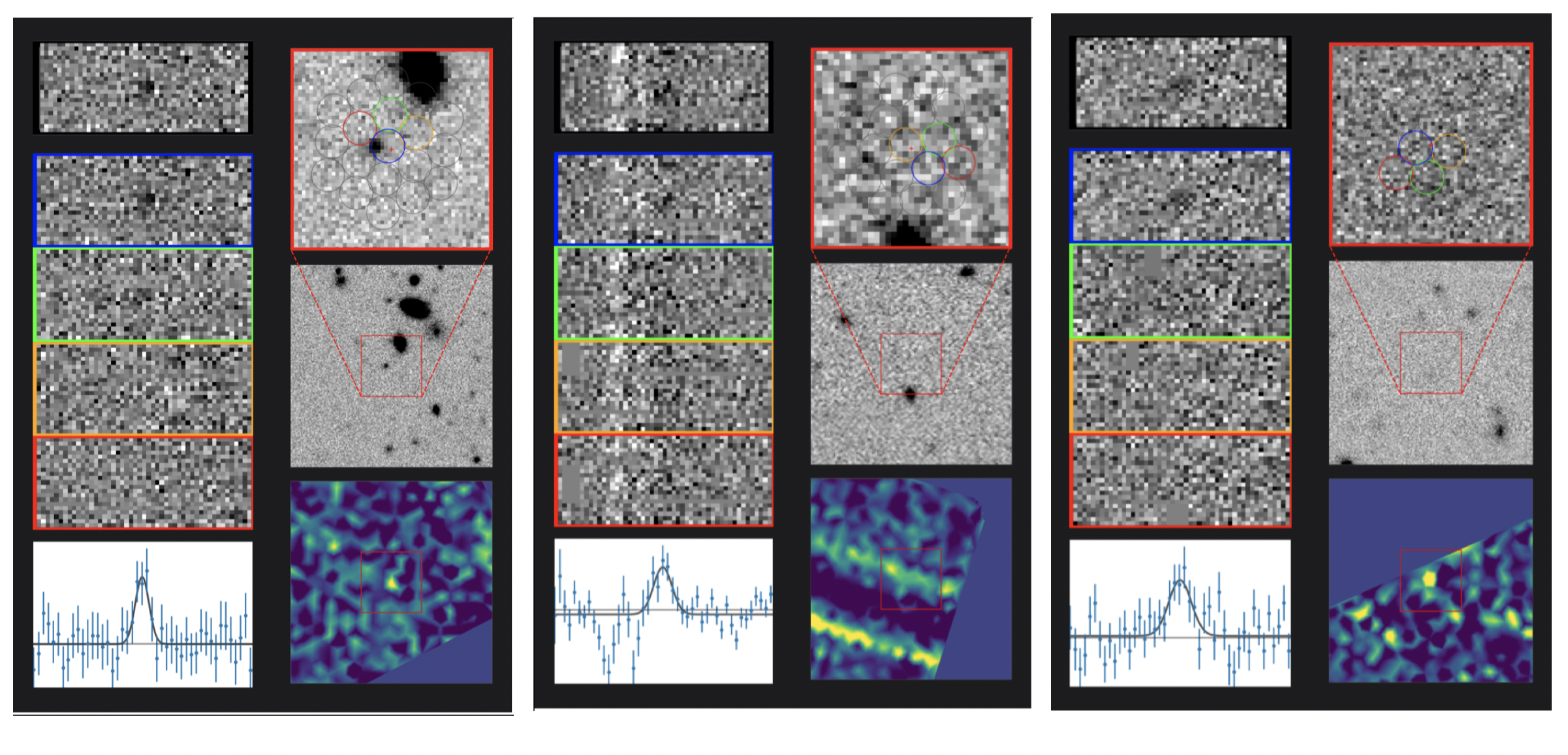}
  \caption{Above, we have examples of the ``mini's" from Dark Energy Explorers ``Fishing for Signal in a Sea of Noise" workflow. From left to right: keep (real galaxy/emission line), throwback (bad detection), and a tricky case that might need more information.}
  \label{fig:KeepThrowbackTossup}
      \end{center}
\end{figure*} 

HETDEX is a untargeted survey, which refers to observations that are conducted without a specific predetermined focus on particular objects, eliminating selection bias, yet creating difficulties when classifying. In this paper we discuss an innovative approach to a data pipeline which will allow accurate labeling efficiently. HETDEX uses the LAEs as a cosmological tracer. Thus, our objects of interest are the LAE galaxies, which must be sorted and selected from the whole sample of objects collected. Just as important in selecting these galaxies is the removal of artifacts, false positives, \OII emitting galaxies, meteor trails, and other emission-like features. LAEs have been detected over an extensive redshift range, and their redshifted 1215.67Å line can easily be detected using low-resolution spectroscopy or narrow-band imaging \citep{Gronwall_2007}. The large-scale clustering of the LAEs will allow us to see the effects of dark energy on the universe and yield the cosmological parameters of scientific interest.

We must differentiate between false positives and galaxy misclassifications since these two errors affect the HETDEX correlation analysis differently. For false positives, we are referring to noise or pixel defects that are confused as an apparent emission line. The misidentification of galaxies is a larger issue, especially when [O II] emitters are designated as LAEs. In this case, the clustering signal of the [O II] galaxies will leave an imprint on the clustering of the LAEs. Therefore, this demonstrates the importance of a robust, clean catalog and will showcase how Dark Energy Explorers and machine learning are powerful techniques for addressing this issue. Here we establish a method of how visual vetting has the ability to significantly improve on the HETDEX algorithms.

Zooniverse\footnote{\url{https://www.zooniverse.org}}, the world's largest participatory science platform, allows us to make progress on visually vetting a large subset of the HETDEX dataset that can be applied to the full sample. With the whole HETDEX catalog classified, many millions of sources labeled, it would enable significant improvements towards the measurements of the HETDEX cosmological parameters \citep{House_2023}. 

We show here how to use participatory science and machine learning to classify 1.2 million of HETDEX sources. In Section~~\ref{sec:citsci}, we discuss participatory science and the success of \textit{Dark Energy Explorers} to explore various HETDEX regimes and statistics we use for the classification. Section~~\ref{subsec:machine learing} focuses on the machine learning algorithm and how we incorporate millions of sources with visually vetted sources. Finally, section~~\ref{sec:incorporating} shows how we utilize the algorithm and  \textit{Dark Energy Explorers} in the creation of a data pipeline for the current HETDEX sample.

\section{Dark Energy Explorers: A Participatory Science Campaign}
\label{sec:citsci}

Participatory Science has continued to grow with organizations like Zooniverse, SciStarter\footnote{\url{https://scistarter.org/}}, and CitizenScience.gov\footnote{\url{https://www.citizenscience.gov/}}. To date, Zooniverse alone has garnered millions of participants and has launched close to 500 projects. Demonstrating the clear demand for visual classification in research, and it has been shown that combining human and machine classifications can efficiently produce results superior to those of either one alone \citep{trouille_citizen_2019-1}. This combination of techniques is what we aim to accomplish with \textit{Dark Energy Explorers}.

\subsection{Developing Dark Energy Explorers} 

We launched \textit{Dark Energy Explorers} in February 2021 and the project has had an impact of roughly six million classifications in the project’s lifetime so far. \textit{Zooniverse} is the host platform that has cultivated a participatory science community and has launched hundreds of projects. Once on Zooniverse, app or website, participants must login or create a new account to save their classifications. As an official NASA participatory Science project we can also be found under the Citizen Science Section of NASA.gov.\footnote{\url{https://science.nasa.gov/citizen-science/dark-energy-explorers/}}  Once on the \textit{Dark Energy Explorers} home page, choose the active workflow, “Fishing for Signal in a Sea of Noise," which will prompt a tutorial if you are a new visitor. The tutorial demonstrates how to classify the HETDEX data (discussed in more detail in \citealt{House_2023}).

When creating this tool to be used by the general public, it was essential to simplify the classification process into digestible, jargon-free tasks. The tutorial and field guide serve as the mechanism to do just that. The tutorial walks participants through what to look for in the HETDEX data in just a few easy-to-understand steps. The primary criteria users consider have not changed from the prior work in \citep{House_2023}. As a reminder, for the “Fishing for Signal in a Sea of Noise" workflow participants classify sources based on:

\begin{enumerate}
    \item The quality of the data collected,  
    \item The strength of the emission line, and
    \item The appearance of the emission line in at least one or more of the fiber spectrum.
\end{enumerate}

Once trained, the participants are provided deidentified, processed data in the form of user friendly imaging. The volunteers must select between two options, Keep this Galaxy, or Throwback.  The field guide is a shortened tutorial allowing quick access to classification reminders. See Figure ~\ref{fig:KeepThrowbackTossup} for a comparison. Figure ~\ref{fig:KeepThrowbackTossup} shows an example of three sources and what they would look like to the participants on \textit{Dark Energy Explorers}. In summary, these show a combination of the various types of data we collect with the HET – fiber cutouts, imaging, flux map, and emission line. See \cite{House_2023} for a detailed description of each \textit{Dark Energy Explorers} input image. On the left is an example of a real galaxy or LAE; in the middle is a source that is an artifact; on the right is a source that is possibly real is a possibility to be real, but the HETDEX team would need to explore more information to decide. This is one of the key advantages of \textit{Dark Energy Explorers} because it allows for a triage of sources that lie in this unclear regime, and the team can quickly identify areas that need a closer look; rather than sifting through all the sources ourselves, this technique creates a smaller, more manageable small subset.  

\begin{figure*}[ptb]
   \begin{center}
         \includegraphics[scale=0.53]{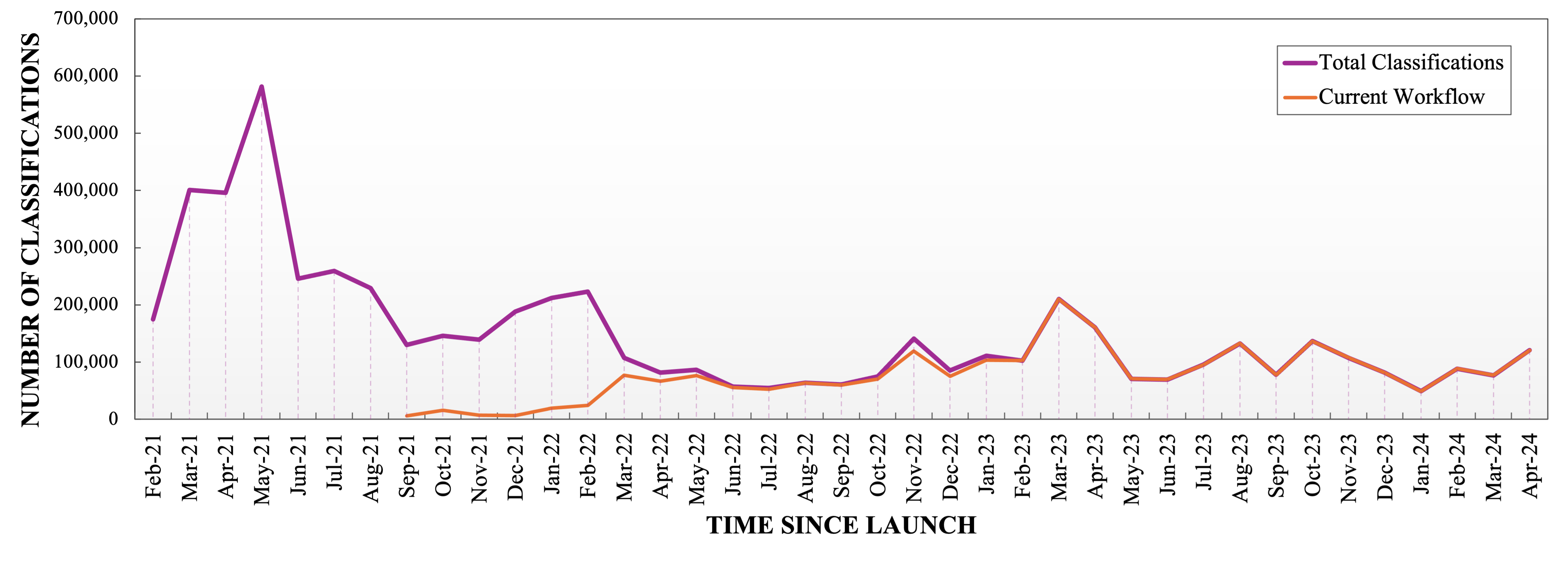}
  \caption{The total classifications, where each source is classified by a minimum of 10 volunteers, collected on \textit{Dark Energy Explorers} since launch. The orange line represents the current active workflow classifications ``Fishing for a Signal in a Sea of Noise'' and the purple represents the total classifications by month.} 
  \label{fig:classifications}
      \end{center}
\end{figure*} 

From \cite{House_2023, erincooper, Gebhardt21, Davis2022} it is essential to explore lower SN regimes to maximize the number of LAE candidates that will lead to a more precise HETDEX cosmology. The previous work began with SN $>$ 6 sources when using \textit{Dark Energy Explorers}, but this work has shown we can progress to lower SN regimes while still remaining accurate with visual vetting. So far, $\approx$ 190,000 LAE candidates have been classified by the \textit{Dark Energy Explorers} down to a SN ratio of 4.8. This includes HETDEX Data Release 2,3,4, and the COSMOS field.

\subsection{Using Dark Energy Explorers as an Education and Public Engagement Tool}

The research results and improvements to HETDEX would not be possible without the classifications from our \textit{Dark Energy Explorers} participants. While the primary goal of \textit{Dark Energy Explorers}is to improve the accuracy of the HETDEX cosmology, it has quickly demonstrated the impact it has as an extraordinary educational tool. Collaboration with McDonald Observatory, where the Hobby-Eberly Telescope is based, has allowed us to grow our volunteer base of \textit{Dark Energy Explorers} while also allowing unique opportunities at the Observatory. We currently have an exhibit at the Hobby-Eberly Telescope (HET)visitors center, which allows a rare experience of visiting a telescope and then being able to classify the data as an amateur researcher. In addition, McDonald Observatory has assisted in developing worksheets and videos for educators to use in traditional classrooms, libraries, and museums. We have continued engagement with our volunteers through Teleconferencing/Zoom nights, our \textit{Dark Energy Explorers} YouTube Channel\footnote{\url{https://www.youtube.com/@DarkEnergyExplorers}}, design competitions, and blog posts. The following section discusses the overall results of implementing these public engagement efforts, and we hope it continues to be a rewarding educational outlet for our dedicated participants. \\

\subsection{Results and Impact of Dark Energy Explorers}
Our overall goal is to vet all HETDEX sources visually. By the end of HETDEX, we will have nearly 10 million sources; this is effectively impossible to do within the team. The only possibility is to include many participatory scientists in classifying the sources. We have been doing this through the \textit{Dark Energy Explorers} since February 2021. \textit{Dark Energy Explorers} has already proven incredibly successful at accomplishing this goal, with roughly six million classifications in the project's lifetime. For the current workflow, that is approximatley 190,000 unique spectra that are identified by a minimum of 10 individuals. Figure ~\ref{fig:classifications} shows our classifications as a function of time, where the purple line demonstrates the total number of classifications each month since \textit{Dark Energy Explorers} launched with the first workflow, ‘Nearby VS Distant,’ which has since been retired \citep{House_2023}. 

The current workflow `Fishing for Galaxies in a Sea of Noise' has 2.1 million classifications resulting in 190,000 completed LAE candidates, which are vetted by at least 10 different participants, giving us the confidence to rely on this average \citep{House_2023}. Figure ~\ref{fig:classifications} represents the current workflow in orange and the total classifications since its launch in September of 2021. These millions of classifications have been done by approximately 17,000 volunteers that represent over 159 countries all over the world, with the top three being in the United States, United Kingdom, and India. The \textit{Dark Energy Explorers} average 153,000 classifications per month and we hope to increase this through outreach and engagement efforts to reach our goal of the entire HETDEX catalog being classified.

\begin{figure*}[ptb]
   \begin{center}
         \includegraphics[scale=0.58]{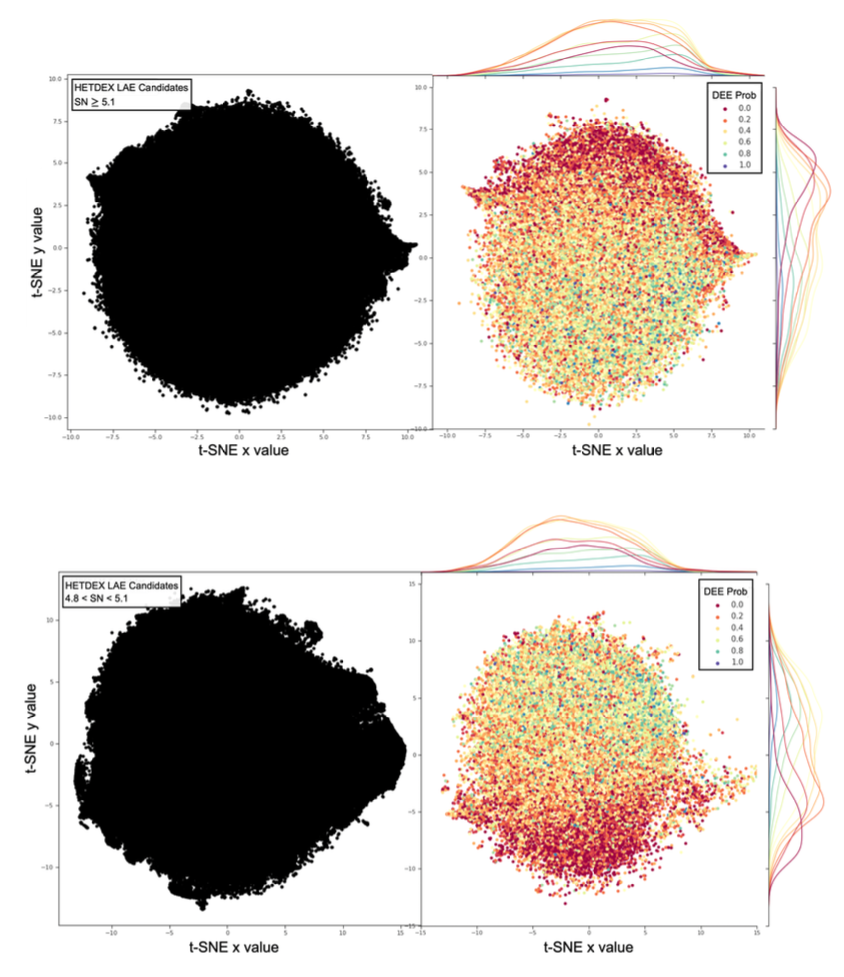}
  \caption{Above shows the results of the t-SNE machine learning algorithm, each with the same hyper-parameters and each row has the same t-SNE projection. On the left side, the plots shown in black, are $\simeq$ 1.2 million HETDEX LAE Candidates. On the right side, the colored plots shown, are a result of the $\simeq$ 190,000 LAE candidates that have been labeled by \textit{Dark Energy Explorers} and assigned a DEE\_probability. The top two plots show the LAE candidates above SN ratio $\geq$ 5.1 and the two bottom plots show the LAE Candidates with  $4.8<$ SN $<5.1$. Note: The orientation of the t-SNE axes relative to the data points has no inherent meaning or significance beyond the visualization itself. This is simply a characteristic of the dimensionality reduction algorithm.}
  \label{fig:tsne_dee}
      \end{center}
\end{figure*}

\begin{figure}[ptb]
         \includegraphics[scale=0.95]{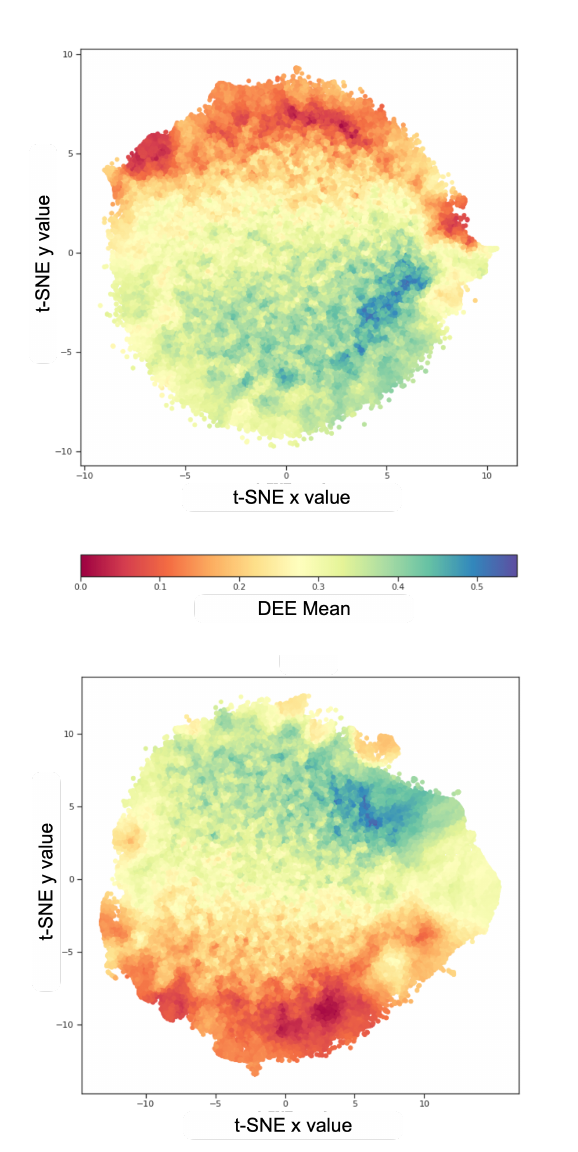}
  \caption{Above shows the results of the t-SNE machine learning algorithm, each with the same hyper-parameters. The colors represent the DEE\_mean which was developed using the nearest neighbors method. Note: The orientation of the t-SNE axes relative to the data points has no inherent meaning or significance beyond the visualization itself. This is simply a characteristic of the dimensionality reduction algorithm.}
  \label{fig:tsne_nn}
\end{figure} 

\section{Visualization in Machine Learning }
\label{subsec:machine learing}

While machine learning (ML) has been shown to be effective many aspects of research are still met with obstacles that require human verification \citep{ML_humans}. This combination of machines and humans, we argue, is a way to make a large problem more efficient and accurate. In particular, it provides necessary insight to the interpretation of the data, in the case of HETDEX, the high-dimensional spectral elements \citep{crowston_blending_2017, Torney}. Other participatory science projects have led the way in imaging classification with both ML and human vetting, yet we discuss how this can be done with a spectroscopic inputs \citep{Masters_2019, simmons_galaxy_2016}.

For the \textit{Dark Energy Explorers} project utilizing machine learning has proven to be most useful for identifying artifacts. We have done this through the `Fishing for Signal in a Sea of Noise' workflow discussed in Section ~~\ref{sec:citsci}. Using the same DEE\_probability from prior work, we assign each classification a value of 1.0 meaning an object is a real LAE detection or a 0.0 which identifies a false positive or artifact. Aggregating over a minimum of ten \textit{Dark Energy Explorers} participants the values are then averaged to generate a DEE\_probability. Once those labels have been acquired and the DEE\_probability determined for each visually vetted source we can apply a machine learning algorithm.

\subsection{Algorithm: t-SNE}

The machine learning algorithm discussed and used in this work is known as t-distributed stochastic neighbor embedding or t-SNE. t-SNE serves as a statistical method for visualizing high-dimensional data by giving each data point a location in a two or three-dimensional map \citep{vandermaaten08, Maaten2015}. For the plethora of high-dimensional data that we get from HETDEX, there will be millions of elements with 1036 dimensions, this will result in billions of spectral elements that t-SNE can handle exceptionally well. 

Using unsupervised machine learning, specifically t-SNE, for large data sets has proven effective when the parameters are optimized accordingly \citep{belkina_2019_ML}. t-SNE has a cost function that with different initialization, like the HETDEX spectra and algorithm parameters, we can get different results. The results depend on the random seed, the data input and the hyper parameters chosen, despite this, t-SNE visualisations can be effective in grouping together sources with similar spectra, especially artifacts (e.g. House et al 2023). 

The results are not reproducible, but when using tuned hyperparameters, it will keep the global aspects of the data, and in the case of cleaning the catalog of artifacts, like in HETDEX, this works well. T-SNE and many dimensionality reduction algorithms can be tricky to interpret \citep{vandermaaten08} and the data from Dark Energy Explorers provides context to interpret this data in a scientifically meaningful way. First, we set the dimensionality reduction to reduce from 50 to 2. Importantly, the perplexity parameter has been shown to give the best results with values of 5-50 \citep{Maaten2015}. Therefore the perplexity found to be optimized at 30 and combining this with an initial iteration parameter of 1000 ensured the algorithm reached a stable configuration. Utilizing the {\tt scikit-learn} Python package, the left side of Figure ~\ref{fig:tsne_dee} displays the results of HETDEX LAE candidates in black after using t-SNE \citep{scikitlearnPedregosa11}. 

\subsection{Input Selection Criteria}
 
To better differentiate our sources of interest, the Lyman-alpha emitting galaxies, we select for the Lyman-alpha emission line in our HETDEX catalog. The spectral elements of these LAE candidates are then used to cut around the emission line of the 1D spectra. We select for 50 angstroms on either side of the Lyman-$\alpha$ emission. This results in a 100-angstrom cut, in which HETDEX captures flux in 2-angstrom bins, resulting in 50 dimensions used for the t-SNE input for each LAE candidate. We further distribute into two sub-samples by signal-to-noise (SN) ratio to downsize our sample of 1.2 million sources. The result of the t-SNE runs can be shown in Figure ~\ref{fig:tsne_dee}. Shown at the top of Figure ~\ref{fig:tsne_dee} is the SN ratio range $\geq$ 5.1 with $\sim$ 600,000 sources. Similarly, shown at the bottom of Figure ~\ref{fig:tsne_dee} is the range of 5.1 $>$ SN $>$ 4.8 with another $\sim$ 600,000 sources for a total of 1.2 M LAE candidate detections. Our labels from \textit{Dark Energy Explorers} that are fed into t-SNE include the low SN sources (6 $>$ SN $>$ 4.8) to ensure the algorithm can train on a similar low-SN sample and eliminate bias. The full SN range from \textit{Dark Energy Explorers} is then visualized in color, labeled, and shown in Figure ~\ref{fig:tsne_dee} for each SN bin, respectively. 

\section{Incorporating the Dark Energy Explorers Results with Machine Learning}
\label{sec:incorporating}

As large astronomical survey's grow, so do the artifacts and contamination. These artifacts are either found manually \citep{erincooper} or with artificial intelligence \citep{Gebhardt21}. While artificial intelligence can be useful, the elements from HETDEX are better originally identified by visual vetting. Here we discuss how we analyze the visual classifications from the participants of \textit{Dark Energy Explorers} and then use the classifications to expand to the full 1.2 million LAE candidates.

\subsection{Analysis of Visual Vetting Statistics}

Following the original work, we focus on the false positives. Using the same methods from prior work, a DEE\_probability of 1.0 means an object is a real LAE detection, and a probability of 0.0 identifies a false positive or artifact. Therefore, every source is identified by a minimum of ten \textit{Dark Energy Explorers} participants and averaged together to get a DEE\_probability. 

Previously, in \cite{House_2023}, a DEE\_probability $\le 0.3$ results in 92\% accuracy across SN. In addition, \cite{House_2023}, demonstrated a \textit{Dark Energy Explorers} probability of below 0.1 gave $98\% $ agreement, which allowed for very efficient and accurate identification of false positives. All of these sources from the pilot sample were cross-examined by members of the HETDEX team which resulted in the development of the DEE\_probability accuracy. Given this high accuracy for identifying false positives we could confidently move forward with applying these results to a broader full sample with machine learning. Those are the results explained here. 


    \begin{figure*}[ptb]
    \begin{center}
         \includegraphics[scale=0.7]{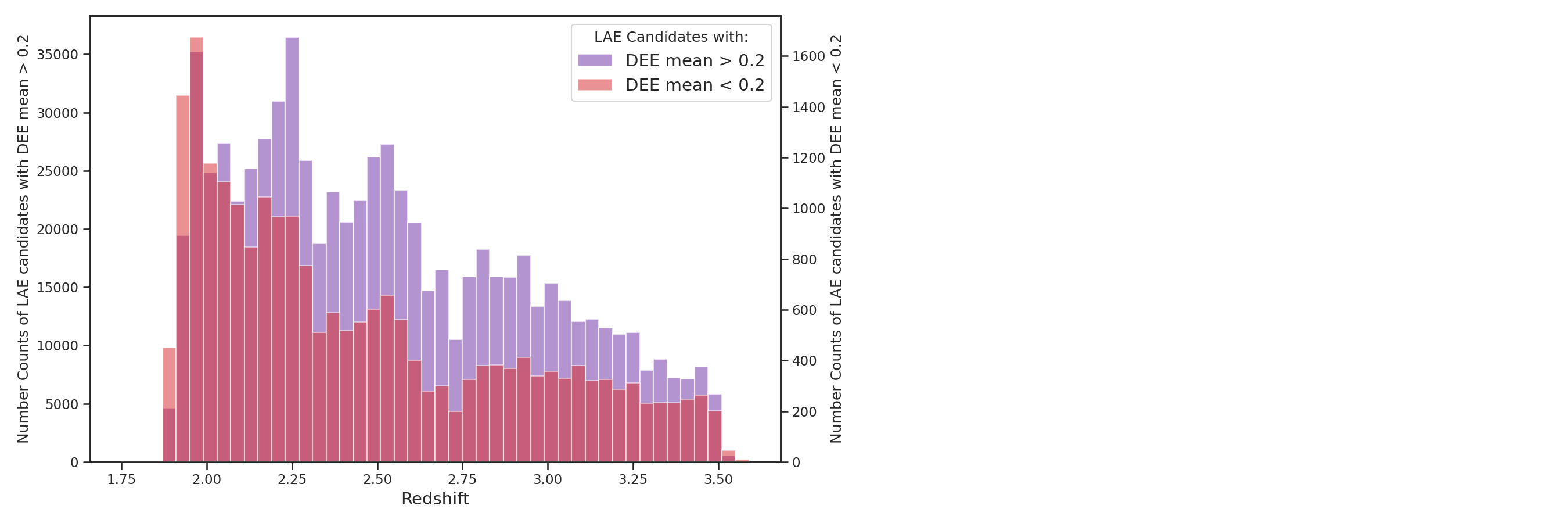}
  \caption{The redshift distribution of the 1.2 million HETDEX LAE candidates with the cut on DEE\_mean. DEE\_mean is created using nearest neighbors technique. The histogram in pink identifies the false positives with a DEE\_mean $<$ 0.2 and the purple represents a DEE\_mean $>$ 0.2, each cut is scaled for comparison, differing by approximately a factor of 20.} 
  \label{fig:redshift}
      \end{center}
\end{figure*} 

\subsection{Application to the Full HETDEX LAE Catalog using Nearest Neighbors technique}

Following the visual vetting analysis, our next goal is to apply this approach to the full HETDEX LAE catalog as opposed to just a subset. Figure~~\ref{fig:tsne_dee} shows the distribution in t-SNE space for the full sample and for those classified by \textit{Dark Energy Explorers}. The top-left panel shows $\sim$600,000 sources for S/N$>5.1$. The bottom-left panels shows the $\sim$600,000 sources that have $4.8<$ SN $<5.1$. The right panels are the corresponding t-SNE distributions for those sources that have a DEE\_probability, with the points colored by their probability.

For each source in the panels on the left, we find the 50 nearest sources in t-SNE space that have a DEE\_probability. The average of these 50 is then applied to each of the sources, which we call the DEE\_mean. Following the same logic as the DEE\_probability, the DEE\_mean ranges from 0 to 1.0, from lowest to highest probability of being a real galaxy detection. The DEE\_mean becomes an additional statistical method for interpreting the t-SNE analysis. 

Figure~\ref{fig:tsne_nn} shows the same t-SNE distribution as in Figure~\ref{fig:tsne_dee}, except we now color the points by their DEE\_mean, generated from the nearest neighbors method. It is clear that there is strong clustering for both sources with high and low probabilities of being real. As stated, we find that the \textit{Dark Energy Explorers} are best at identifying false positive due to instrumental and reduction artifacts, and we use the DEE\_mean to further remove these sources. Thus, the red points in Figure~\ref{fig:tsne_nn} are considered false detections and will be removed from subsequent analysis. We find a cut of DEE\_mean$<0.2$ provides an adequate cut.


\section{Overall Implications to HETDEX and the Cosmology}

Similar to \cite{House_2023}, we rely on visually vetting from the HETDEX team in order to determine accuracy for both DEE\_probability and DEE\_mean. We previously determined an accuracy of 92\% for DEE\_probability $<0.3$, implying that when 7 out of 10 \textit{Dark Energy Explorers} call a source false, they are correct 92\% of the time. 

For DEE\_mean, the HETDEX team visually vets about 300 sources at random, and we find that for DEE\_mean $<0.1$ we agree 94\% of the time and DEE\_mean$<0.2$ we agree 91\% of the time. We use this cut as a further technique to remove false positives from the HETDEX dataset. In this way, we are able to use the participatory scientists for the full sample. Eventually, we plan to visually vet all sources, without having to use a nearest neighbor approach. 

For the current dataset of 1.2 million sources, we have 62,000 with DEE\_mean $<0.2$. These will be removed from the sample. While they only represented 5\%, they may have certain aspects that could bias the cosmological analysis and therefore their removal is essential. The most obvious concern is shown in Figure~\ref{fig:redshift}. In this figure, we show the histogram of the full sample and a scaled (roughly 20 fold) histogram of the sources that we remove based on the DEE\_mean. There is clearly a bias towards low redshifts for the sources that are being removed. The point is that there appears to be an increase in the contamination rate as a function of redshift. The overall contamination remains small, and we will study whether this bias could have implications for the cosmology in future work. 

An important aspect is that we do not necessarily know the truth. For example, the HETDEX team, while having the deepest understanding of the dataset, can make mis-classifications. It is possible that having multiple individuals, as we do in \textit{Dark Energy Explorers}, provide a more robust result. Additionally, using a nearest neighbor approach could be more robust than the individual measures, since it relies on averaging within t-SNE space. In the end, HETDEX will use multiple measures of the source classification to understand the influence on the cosmological analysis. The DEE\_prob and DEE\_mean values and recommendations will be included in the future HETDEX Data Release 3 (HDR3) catalog.

\textit{Dark Energy Explorers} has had an extremely positive impact on our informal science community and on our overall goal to improve the accuracy of HETDEX. In future work we aim to continue both these efforts to reach a HETDEX catalog that has been completely visually classified. This will be roughly a factor of 10 more than what we have now and we will continue to have pave the way for a successful way to use machine learning and participatory science.



\section*{acknowledgments}

The results of the Dark Energy Explorers would not be as robust and useful if not for the care and dedication by the volunteers. We are extremely grateful to their work. It is having a large impact and is motivational.

LH acknowledges support from NSF GRFP DGE 2137420 and NASA 21-CSSFP21-0009. KG acknowledges support from the NSF-2008793 and from NASA 21-CSSFP21-0009.

\textit{Dark Energy Explorers} is recognized as an official NASA Citizen Science partner. This publication uses data generated via the \href{https://www.zooniverse.org}{Zooniverse.org} platform, development of which is funded by generous support, including a Global Impact Award from Google, and by a grant from the Alfred P. Sloan Foundation. 

HETDEX  is led by the University of Texas at Austin McDonald Observatory and Department of Astronomy with participation from the Ludwig-Maximilians-Universit\"at M\"unchen, Max-Planck-Institut f\"ur Extraterrestrische Physik (MPE), Leibniz-Institut f\"ur Astrophysik Potsdam (AIP), Texas A\&M University, The Pennsylvania State University, Institut f\"ur Astrophysik G\"ottingen, The University of Oxford, Max-Planck-Institut f\"ur Astrophysik (MPA), The University of Tokyo, and Missouri University of Science and Technology. In addition to Institutional support, HETDEX is funded by the National Science Foundation (grant AST-0926815), the State of Texas, the US Air Force (AFRL FA9451-04-2-0355), and generous support from private individuals and foundations.

The Hobby-Eberly Telescope (HET) is a joint project of the University of Texas at Austin, the Pennsylvania State University, Ludwig-Maximilians-Universit\"at M\"unchen, and Georg-August-Universit\"at G\"ottingen. The HET is named in honor of its principal benefactors, William P. Hobby and Robert E. Eberly. The Institute for Gravitation and the Cosmos is supported by the Eberly College of Science and the Office of the Senior Vice President for Research at the Pennsylvania State University.

The authors acknowledge the Texas Advanced Computing Center (TACC) \footnote{\url{http://www.tacc.utexas.edu}} at The University of Texas at Austin for providing high performance computing, visualization, and storage resources that have contributed to the research results reported within this paper. This research made use of {\tt scikit-learn}, an open-source Python package for machine learning.

\clearpage

\bibliographystyle{aasjournal}
\bibliography{paper}

\begin{thebibliography}{}
\expandafter\ifx\csname natexlab\endcsname\relax\def\natexlab#1{#1}\fi
\providecommand{\url}[1]{\href{#1}{#1}}
\providecommand{\dodoi}[1]{doi:~\href{http://doi.org/#1}{\nolinkurl{#1}}}
\providecommand{\doeprint}[1]{\href{http://ascl.net/#1}{\nolinkurl{http://ascl.net/#1}}}
\providecommand{\doarXiv}[1]{\href{https://arxiv.org/abs/#1}{\nolinkurl{https://arxiv.org/abs/#1}}}

\bibitem[{Amershi {et~al.}(2014)Amershi, Cakmak, Knox, \& Kulesza}]{ML_humans}
Amershi, S., Cakmak, M., Knox, W.~B., \& Kulesza, T. 2014, AI Magazine, 35, 105, \dodoi{10.1609/aimag.v35i4.2513}

\bibitem[{Belkina {et~al.}(2019)Belkina, Ciccolella, Anno, Halpert, Spidlen, \& Snyder-Cappione}]{belkina_2019_ML}
Belkina, A.~C., Ciccolella, C.~O., Anno, R., {et~al.} 2019, Nat Commun, 10, 5415, \dodoi{10.1038/s41467-019-13055-y}

\bibitem[{Crowston {et~al.}(2017)Crowston, Àsterlund, \& Lee}]{crowston_blending_2017}
Crowston, K., Àsterlund, C., \& Lee, T.~K. 2017, Hawaii International Conference Preceedings on System Sciences, \dodoi{10.24251/HICSS.2017.009}

\bibitem[{{Davis} {et~al.}(2023){Davis}, {Gebhardt}, {Mentuch Cooper}, {Ciardullo}, {Fabricius}, {Farrow}, {Feldmeier}, {Finkelstein}, {Gawiser}, {Gronwall}, {Hill}, {Hopp}, {House}, {Jeong}, {Kollatschny}, {Komatsu}, {Landriau}, {Liu}, {Saito}, {Tuttle}, {Wold}, {Zeimann}, \& {Zhang}}]{Davis2022}
{Davis}, D., {Gebhardt}, K., {Mentuch Cooper}, E., {et~al.} 2023, arXiv e-prints, arXiv:2301.01799.
\newblock \doarXiv{2301.01799}

\bibitem[{Farrow {et~al.}(2021)Farrow, S{\'{a} }nchez, Ciardullo, Cooper, Davis, Fabricius, Gawiser, Gebhardt, Gebhardt, Hill, Jeong, Komatsu, Landriau, Liu, Saito, Snigula, \& Wold}]{Farrow2021}
Farrow, D.~J., S{\'{a} }nchez, A.~G., Ciardullo, R., {et~al.} 2021, Monthly Notices of the Royal Astronomical Society, 507, 3187, \dodoi{10.1093/mnras/stab1986}

\bibitem[{Fluke \& Jacobs(2020)}]{ML_overview_astro}
Fluke, C.~J., \& Jacobs, C. 2020, WIREs Data Mining and Knowledge Discovery, 10, e1349, \dodoi{https://doi.org/10.1002/widm.1349}

\bibitem[{{Gebhardt} {et~al.}(2021){Gebhardt}, {Mentuch Cooper}, {Ciardullo}, {Acquaviva}, {Bender}, {Bowman}, {Castanheira}, {Dalton}, {Davis}, {de Jong}, {DePoy}, {Devarakonda}, {Dongsheng}, {Drory}, {Fabricius}, {Farrow}, {Feldmeier}, {Finkelstein}, {Froning}, {Gawiser}, {Gronwall}, {Herold}, {Hill}, {Hopp}, {House}, {Janowiecki}, {Jarvis}, {Jeong}, {Jogee}, {Kakuma}, {Kelz}, {Kollatschny}, {Komatsu}, {Krumpe}, {Landriau}, {Liu}, {Niemeyer}, {MacQueen}, {Marshall}, {Mawatari}, {McLinden}, {Mukae}, {Nagaraj}, {Ono}, {Ouchi}, {Papovich}, {Sakai}, {Saito}, {Schneider}, {Schulze}, {Shanmugasundararaj}, {Shetrone}, {Sneden}, {Snigula}, {Steinmetz}, {Thomas}, {Thomas}, {Tuttle}, {Urrutia}, {Wisotzki}, {Wold}, {Zeimann}, \& {Zhang}}]{Gebhardt21}
{Gebhardt}, K., {Mentuch Cooper}, E., {Ciardullo}, R., {et~al.} 2021, \apj, 923, 217, \dodoi{10.3847/1538-4357/ac2e03}

\bibitem[{Gronwall {et~al.}(2007)Gronwall, Ciardullo, Hickey, Gawiser, Feldmeier, van Dokkum, Urry, Herrera, Lehmer, Infante, Orsi, Marchesini, Blanc, Francke, Lira, \& Treister}]{Gronwall_2007}
Gronwall, C., Ciardullo, R., Hickey, T., {et~al.} 2007, The Astrophysical Journal, 667, 79, \dodoi{10.1086/520324}

\bibitem[{Hill {et~al.}(2021)Hill, Lee, MacQueen, Kelz, Drory, Vattiat, Good, Ramsey, Kriel, Peterson, DePoy, Gebhardt, Marshall, Tuttle, Bauer, Chonis, Fabricius, Froning, Häuser, Indahl, Jahn, Landriau, Leck, Montesano, Prochaska, Snigula, Zeimann, Bryant, Damm, Fowler, Janowiecki, Martin, Mrozinski, Odewahn, Rostopchin, Shetrone, Spencer, Cooper, Armandroff, Bender, Dalton, Hopp, Komatsu, Nicklas, Ramsey, Roth, Schneider, Sneden, \& Steinmetz}]{Hill_2021}
Hill, G.~J., Lee, H., MacQueen, P.~J., {et~al.} 2021, The Astronomical Journal, 162, 298, \dodoi{10.3847/1538-3881/ac2c02}

\bibitem[{{House} {et~al.}(2023){House}, {Gebhardt}, {Finkelstein}, {Cooper}, {Davis}, {Ciardullo}, {Farrow}, {Finkelstein}, {Gronwall}, {Jeong}, {Johnson}, {Liu}, {Thomas}, \& {Zeimann}}]{House_2023}
{House}, L.~R., {Gebhardt}, K., {Finkelstein}, K., {et~al.} 2023, \apj, 950, 82, \dodoi{10.3847/1538-4357/accdd0}

\bibitem[{Kimura \& Kinchy(2016)}]{cs_ps2}
Kimura, A., \& Kinchy, A. 2016, Engaging Science, Technology, and Society, 2, 331, \dodoi{10.17351/ests2016.99}

\bibitem[{Leung {et~al.}(2017)Leung, Acquaviva, Gawiser, Ciardullo, Komatsu, Malz, Zeimann, Bridge, Drory, Feldmeier, Finkelstein, Gebhardt, Gronwall, Hagen, Hill, \& Schneider}]{Leung2017}
Leung, A.~S., Acquaviva, V., Gawiser, E., {et~al.} 2017, The Astrophysical Journal, 843, 130, \dodoi{10.3847/1538-4357/aa71af}

\bibitem[{Masters(2019)}]{Masters_2019}
Masters, K.~L. 2019, Proceedings of the International Astronomical Union, 14, 205–212, \dodoi{10.1017/S1743921319008615}

\bibitem[{{Mentuch Cooper} {et~al.}(2023){Mentuch Cooper}, {Gebhardt}, {Davis}, {Farrow}, {Liu}, {Zeimann}, {Ciardullo}, {Feldmeier}, {Drory}, {Jeong}, {Benda}, {Bowman}, {Boylan-Kolchin}, {Chavez Ortiz}, {Debski}, {Dentler}, {Fabricius}, {Farooq}, {Finkelstein}, {Gawiser}, {Gronwall}, {Hill}, {Hopp}, {House}, {Janowiecki}, {Khoraminezhad}, {Kollatschny}, {Komatsu}, {Landriau}, {Lujan Niemeyer}, {Lee}, {MacQueen}, {Mawatari}, {McKay}, {Ouchi}, {Poppe}, {Saito}, {Schneider}, {Snigula}, {Thomas}, {Tuttle}, {Urrutia}, {Weiss}, {Wisotzki}, {Zhang}, \& {The HETDEX collaboration}}]{erincooper}
{Mentuch Cooper}, E., {Gebhardt}, K., {Davis}, D., {et~al.} 2023, arXiv e-prints, arXiv:2301.01826.
\newblock \doarXiv{2301.01826}

\bibitem[{Pedregosa {et~al.}(2012)Pedregosa, Varoquaux, Gramfort, Michel, Thirion, Grisel, Blondel, Prettenhofer, Weiss, Dubourg, VanderPlas, Passos, Cournapeau, Brucher, Perrot, \& Duchesnay}]{scikitlearnPedregosa11}
Pedregosa, F., Varoquaux, G., Gramfort, A., {et~al.} 2012, CoRR, abs/1201.0490

\bibitem[{Simmons {et~al.}(2016)Simmons, Lintott, Willett, Masters, Kartaltepe, Häußler, Kaviraj, Krawczyk, Kruk, {McIntosh}, Smethurst, Nichol, Scarlata, Schawinski, Conselice, Almaini, Ferguson, Fortson, Hartley, Kocevski, Koekemoer, Mortlock, Newman, Bamford, Grogin, Lucas, Hathi, {McGrath}, Peth, Pforr, Rizer, Wuyts, Barro, Bell, Castellano, Dahlen, Dekel, Ownsworth, Faber, Finkelstein, Fontana, Galametz, Grützbauch, Koo, Lotz, Mobasher, Mozena, Salvato, \& Wiklind}]{simmons_galaxy_2016}
Simmons, B.~D., Lintott, C., Willett, K.~W., {et~al.} 2016, Monthly Notices of the Royal Astronomical Society, 464, 4420, \dodoi{10.1093/mnras/stw2587}

\bibitem[{Strasser {et~al.}(2019)Strasser, Bruno, Baudry, Jérôme, Mahr, Dana, Sanchez, Gabriela, Tancoigne, \& Élise}]{cs_ps}
Strasser, Bruno, Baudry, {et~al.} 2019, Science and Technology Studies. Special Issue: Many Modes of Citizen Science, 32, 52–76, \dodoi{10.23987/sts.60425}

\bibitem[{Torney {et~al.}(2019)Torney, Lloyd-Jones, Chevallier, Moyer, Maliti, Mwita, Kohi, \& Hopcraft}]{Torney}
Torney, C.~J., Lloyd-Jones, D.~J., Chevallier, M., {et~al.} 2019, Methods in Ecology and Evolution, 10, 779, \dodoi{https://doi.org/10.1111/2041-210X.13165}

\bibitem[{Trouille {et~al.}(2019)Trouille, Lintott, \& Fortson}]{trouille_citizen_2019-1}
Trouille, L., Lintott, C.~J., \& Fortson, L.~F. 2019, Proceedings of the National Academy of Sciences, 116, 1902, \dodoi{10.1073/pnas.1807190116}

\bibitem[{van~der Maaten(2015)}]{Maaten2015}
van~der Maaten, L. 2015, Journal of Machine Learning Research, 15, 3221

\bibitem[{van~der Maaten \& Hinton(2008)}]{vandermaaten08}
van~der Maaten, L., \& Hinton, G. 2008, Journal of Machine Learning Research, 9, 2579

\bibitem[{Zawacki-Richter {et~al.}(2019)Zawacki-Richter, Marín, Bond, \& Gouverneur}]{zawacki-richter_systematic_2019}
Zawacki-Richter, O., Marín, V.~I., Bond, M., \& Gouverneur, F. 2019, International Journal of Educational Technology in Higher Education, 16, 39, \dodoi{10.1186/s41239-019-0171-0}

\bibitem[{Zevin {et~al.}(2024)Zevin, Jackson, Doctor, Wu, Østerlund, Johnson, Berry, Crowston, Coughlin, Kalogera, Banagiri, Davis, Glanzer, Hao, Katsaggelos, Patane, Sanchez, Smith, Soni, Trouille, Walker, Aerith, Domainko, Baranowski, Niklasch, \& Téglás}]{zevin_gravity_2024}
Zevin, M., Jackson, C.~B., Doctor, Z., {et~al.} 2024, Eur. Phys. J. Plus, 139, 100, \dodoi{10.1140/epjp/s13360-023-04795-4}

\end{thebibliography}

\end{document}